\documentclass[aps,prd,twocolumn,superscriptaddress,showpacs]{revtex4}
\usepackage[dvips]{graphicx}
\usepackage{amsmath,amssymb,times}
\usepackage{braket}

\newcommand{\bequ}{\begin{equation}}
\newcommand{\eequ}{\end{equation}}
\newcommand{\bea}{\begin{eqnarray}}
\newcommand{\eea}{\end{eqnarray}}

\def\ga{\mathrel{\mathpalette\fun >}}
\def\fun#1#2{\lower3.6pt\vbox{\baselineskip0pt\lineskip.9pt
  \ialign{$\mathsurround=0pt#1\hfil##\hfil$\crcr#2\crcr\sim\crcr}}}

\DeclareSymbolFont{boldletters}{OML}{cmm} {b}{it}
\DeclareSymbolFontAlphabet{\mathbit}{boldletters}
\DeclareMathSymbol{\alpha}{\mathalpha}{letters}{"0B}
\DeclareMathSymbol{\beta}{\mathalpha}{letters}{"0C}
\DeclareMathSymbol{\gamma}{\mathalpha}{letters}{"0D}
\DeclareMathSymbol{\delta}{\mathalpha}{letters}{"0E}
\DeclareMathSymbol{\epsilon}{\mathalpha}{letters}{"0F}
\DeclareMathSymbol{\zeta}{\mathalpha}{letters}{"10}
\DeclareMathSymbol{\eta}{\mathalpha}{letters}{"11}
\DeclareMathSymbol{\theta}{\mathalpha}{letters}{"12}
\DeclareMathSymbol{\iota}{\mathalpha}{letters}{"13}
\DeclareMathSymbol{\kappa}{\mathalpha}{letters}{"14}
\DeclareMathSymbol{\lambda}{\mathalpha}{letters}{"15}
\DeclareMathSymbol{\mu}{\mathalpha}{letters}{"16}
\DeclareMathSymbol{\nu}{\mathalpha}{letters}{"17}
\DeclareMathSymbol{\xi}{\mathalpha}{letters}{"18}
\DeclareMathSymbol{\pi}{\mathalpha}{letters}{"19}
\DeclareMathSymbol{\rho}{\mathalpha}{letters}{"1A}
\DeclareMathSymbol{\sigma}{\mathalpha}{letters}{"1B}
\DeclareMathSymbol{\tau}{\mathalpha}{letters}{"1C}
\DeclareMathSymbol{\upsilon}{\mathalpha}{letters}{"1D}
\DeclareMathSymbol{\phi}{\mathalpha}{letters}{"1E}
\DeclareMathSymbol{\chi}{\mathalpha}{letters}{"1F}
\DeclareMathSymbol{\psi}{\mathalpha}{letters}{"20}
\DeclareMathSymbol{\omega}{\mathalpha}{letters}{"21}
\DeclareMathSymbol{\varepsilon}{\mathalpha}{letters}{"22}
\DeclareMathSymbol{\vartheta}{\mathalpha}{letters}{"23}
\DeclareMathSymbol{\varpi}{\mathalpha}{letters}{"24}
\DeclareMathSymbol{\varrho}{\mathalpha}{letters}{"25}
\DeclareMathSymbol{\varsigma}{\mathalpha}{letters}{"26}
\DeclareMathSymbol{\varphi}{\mathalpha}{letters}{"27}
\DeclareMathSymbol{\Gamma}{\mathalpha}{letters}{"00}
\DeclareMathSymbol{\Delta}{\mathalpha}{letters}{"01}
\DeclareMathSymbol{\Theta}{\mathalpha}{letters}{"02}
\DeclareMathSymbol{\Lambda}{\mathalpha}{letters}{"03}
\DeclareMathSymbol{\Xi}{\mathalpha}{letters}{"04}
\DeclareMathSymbol{\Pi}{\mathalpha}{letters}{"05}
\DeclareMathSymbol{\Sigma}{\mathalpha}{letters}{"06}
\DeclareMathSymbol{\Upsilon}{\mathalpha}{letters}{"07}
\DeclareMathSymbol{\Phi}{\mathalpha}{letters}{"08}
\DeclareMathSymbol{\Psi}{\mathalpha}{letters}{"09}
\DeclareMathSymbol{\Omega}{\mathalpha}{letters}{"0A}

\begin{document}
\preprint{SAGA-HE-255-09}
\title{Spontaneous parity and charge-conjugation violations 
at real isospin and imaginary baryon chemical potentials}

\author{Hiroaki Kouno}
\email[]{kounoh@cc.saga-u.ac.jp}
\affiliation{Department of Physics, Saga University,
             Saga 840-8502, Japan}

\author{Mizuho Kishikawa}
\affiliation{E.I.M. Electric Corporation Limited, Fukuoka 807-
0001, Japan}

\author{Takahiro Sasaki}
\email[]{sasaki@phys.kyushu-u.ac.jp}
\affiliation{Department of Physics, Graduate School of Sciences, Kyushu University,
             Fukuoka 812-8581, Japan}

\author{Yuji Sakai}
\email[]{sakai@phys.kyushu-u.ac.jp}
\affiliation{Department of Physics, Graduate School of Sciences, Kyushu University,
             Fukuoka 812-8581, Japan}

\author{Masanobu Yahiro}
\email[]{yahiro@phys.kyushu-u.ac.jp}
\affiliation{Department of Physics, Graduate School of Sciences, Kyushu University,
             Fukuoka 812-8581, Japan}

\date{\today}

\begin{abstract}
The phase structure of two-flavor QCD is investigated 
at real isospin and imaginary quark chemical potentials by using 
the Polyakov-loop extended Nambu--Jona-Lasinio model. 
In the region, parity symmetry is spontaneously broken 
by the pion superfluidity phase transition, 
whereas charge-conjugation symmetry is spontaneously violated 
by the Roberge-Weiss transition. 
The chiral (deconfinement) crossover at 
zero isospin and quark chemical potentials is a remnant 
of the parity (charge-conjugation) violation. 
The interplay between the parity and charge-conjugation violations are 
analyzed, and it is investigated how the interplay is related to 
the correlation between the chiral and deconfinement crossovers 
at zero isospin and quark chemical potentials. 
\end{abstract}

\maketitle

\section{Introduction}
\label{Introduction}

The phase diagram of Quantum Chromodynamics (QCD) is 
the key to understanding not only natural 
phenomena such as compact stars and the early universe but also 
laboratory experiments such as relativistic heavy-ion collisions. 
Quantitative calculations of the phase diagram with 
the first-principle lattice QCD (LQCD) have the well-known sign problem 
when the baryon chemical potential $\mu_{\rm B}$ is real~\cite{Kogut}; 
here $\mu_{\rm B}=3\mu_{\rm q}$ for the quark-number chemical potential 
$\mu_{\rm q}$. 

The grand canonical partition function $Z(\mu_{\rm q} )$ of two-flavor QCD 
can be obtained by 
\begin{eqnarray}
Z(\mu_{\rm q} )=\int {\cal D}U\det{\Delta (\mu_{\rm q})}e^{-S_{\rm G}}, 
\label{eq:Z}
\end{eqnarray}
where $\int {\cal D}U$ denotes the path integral with respect to gauge variable $U$, $S_{\rm G}$ stands for the pure gauge action and 
\begin{eqnarray}
\Delta (\mu_{\rm q}) =D_\nu \gamma_\nu +m_0-\mu_{\rm q}\gamma_4 
\label{eq:Delta}
\end{eqnarray}
with the gamma matrix $\gamma_\mu$ in the Euclidean notation 
and the quark mass $m_0$. 
For simplicity, up and down quarks are assumed to have 
the same mass $m_0$. When $\mu_{\rm q}$ is real, it is easy to verify 
\begin{eqnarray}
\Delta (\mu_{\rm q})^\dagger =-D_\nu\gamma_\nu+m_0-\mu_{\rm q}\gamma_4
=\gamma_5\Delta (-\mu_{\rm q})\gamma_5. 
\label{eq:Delta_dagger}
\end{eqnarray}
This leads to 
\begin{eqnarray}
&\left\{ \det{\Delta (\mu_{\rm q})}\right\}^*
=\det{\left\{\gamma_5\Delta (-\mu_{\rm q})\gamma_5\right\}}
\nonumber\\
&=\det{\Delta (-\mu_{\rm q}) }
\neq \det{\Delta (\mu_{\rm q}) }, 
\label{eq:det_Delta}
\end{eqnarray}
and hence $\det{\Delta (\mu_{\rm q})}$ is complex. 
This causes the sign problem. 
Several approaches have been proposed so far 
to circumvent the difficulty; 
for example, the reweighting method~\cite{Fodor}, 
the Taylor expansion method~\cite{Allton} and 
the analytic continuation from imaginary $\mu_{\rm q}$ 
to real $\mu_{\rm q}$
~\cite{FP,Elia,Chen,D'Elia-iso,Cea,D'Elia-3,FP2010,Nagata,Takaishi}. 
However, those are still far from perfection particularly at 
$\mu_{\rm q}/T \ga 1$, where $T$ is temperature. 

As an approach complementary to LQCD, 
we can consider effective models such as the  
Nambu--Jona-Lasinio (NJL) model
~\cite{NJ1,Klevansky,HK,Buballa,AY,Fujii,Osipov,Kashiwa,FIK,Boer,Boer2,Mizher,He,BCPR,Klimt,Vogl} 
and the Polyakov-loop extended Nambu--Jona-Lasinio (PNJL) model~\cite{Meisinger,Dumitru,Fukushima,Ghosh,Megias,Ratti,Ciminale,Rossner,Hansen,Sasaki,Schaefer,Costa,Kashiwa1,Fu,Abuki,Abuki2,Sakai,Sakai2,Kashiwa5,Kouno,Hell,Sakai3,Bhattacharyya,Fukushima3,Matsumoto,Contrera,Sasaki-T,Sakai5,Gatto,Gatto:2010pt,Zhang,Mukherjee,Fukushima2,McLerran_largeNc,Kouno_CP,Sakai:2011fa,Sakai6,Sakai_R,Kashiwa_MG,Pagura,Kashiwa_NL_IM,Morita,Hell2,Roessner2}. 
The NJL model describes the chiral symmetry breaking, but not 
the confinement mechanism. 
The PNJL model is extended so as 
to treat both the mechanisms~approximately by 
considering the Polyakov loop in addition to the chiral condensate 
as ingredients of the model.

In the NJL-type models, 
the input parameters are usually determined from the 
pion mass and the pion decay constant 
at vacuum ($\mu_{\rm q}=0$ and $T = 0$). 
It is then highly nontrivial whether the models properly predict the dynamics 
of QCD at finite $\mu_{\rm q}$. This should be tested from QCD. 
Such a test is possible at imaginary $\mu_{\rm q}$ and finite 
isospin chemical potential, since LQCD has no sign problem there.

For imaginary quark chemical potential, $\mu_{\rm q}=i\Theta T$, 
one can verify that 
\begin{eqnarray}
\Delta (i\Theta T)^\dagger =-D_\nu\gamma_\nu +m_0+i\Theta T\gamma_4
=\gamma_5\Delta (i\Theta T)\gamma_5. 
\label{eq:Delta_dagger_i}
\end{eqnarray}
This leads to 
\begin{eqnarray}
\left\{\det{\Delta (i\Theta T)}\right\}^*
=\det{\left\{\gamma_5\Delta (i\Theta T)\gamma_5\right\}}
=\det{\Delta (i\Theta T) } ,
\label{eq:det_Delta_i}
\end{eqnarray}
and hence $\det{\Delta (i\Theta T)}$ is real. LQCD has thus 
no sign problem at imaginary $\mu_{\rm q}$.

At imaginary $\mu_{\rm q}$, the QCD partition function  has 
the Roberge-Weiss (RW) periodicity, i.e. the periodicity of $2\pi/3$ 
in $\Theta$. The RW periodicity is a remnant of 
the $\mathbb{Z}_3$ symmetry in the pure gauge limit. 
This periodicity can be reinterpreted as 
the extended $\mathbb{Z}_3$ symmetry~\cite{Sakai_R,Sakai,Sakai2,Kashiwa5}; 
see \eqref{eq:K2}-\eqref{RW-periodicity} 
for the relation between the RW periodicity and 
the extended $\mathbb{Z}_3$ symmetry.
At higher $T$, three $\mathbb{Z}_3$ vacua emerge one by one~\cite{RW} 
as $\Theta$ increases. 
This mechanism guarantees the extended $\mathbb{Z}_3$ symmetry 
(the RW periodicity) at higher $T$. 
The mechanism induces the first-order phase transition 
at $\Theta=(2k+1)\pi/3$, where $k$ is an integer~\cite{RW}. 
This phase transition is called the RW transition. 
The partition function of the PNJL model 
has the extended $\mathbb{Z}_3$ symmetry and hence yields 
the RW periodicity and the RW transition~\cite{Sakai,Kouno,Pagura,Morita,Kashiwa_NL_IM}. 
Eventually, the PNJL model qualitatively reproduces LQCD data~\cite{Sakai}.

LQCD simulations at imaginary $\mu_{\rm q}$ show that  
the chiral and deconfinement crossovers take place 
simultaneously~\cite{FP,Elia,Chen,D'Elia-iso,Cea,D'Elia-3,FP2010,Nagata,Takaishi}. 
Such a strong correlation between the transitions 
is not seen in the original PNJL model\cite{Sakai,Sakai2,Sakai3}. 
We then extended the PNJL model to solve this problem. 
In the new model called the entanglement PNJL (EPNJL) model, the strength of 
the four-quark vertex depends on the Polykov 
loop~\cite{Sakai5}. 
In the EPNJL model, the chiral and deconfinement crossovers 
coincide with each other, so that the EPNJL model yields good agreement 
with LQCD data~\cite{Sakai5}.

The reliability of the PNJL and EPNJL models can be tested at 
finite isospin chemical potential ($\mu_{\rm iso}$), 
since LQCD has no sign problem there~\cite{Son-Stephanov}. 
For later convenience, we use the ``modified" isospin chemical potential 
$\mu_{\rm I}=\mu_{\rm iso}/2$ instead of $\mu_{\rm iso}=\mu_u-\mu_d$, where 
$\mu_u$ and $\mu_d$ are chemical potentials for up and down quarks, 
respectively. The QCD partition function at finite $\mu_{\rm I}$ 
is obtained by 
\begin{eqnarray}
Z(\mu_{\rm I} )=&\int {\cal D}U
\det{\Delta (\mu_u)}\det{\Delta (\mu_d)} 
e^{-S_{\rm G}}
\nonumber\\
=&\int {\cal D}U
\det{\Delta (\mu_{\rm I})}\det{\Delta (-\mu_{\rm I})} 
e^{-S_{\rm G}}. 
\label{eq:Z_iso}
\end{eqnarray}
When $\mu_u$ and $\mu_d$ are pure imaginary, 
both $\det{\Delta (\mu_{\rm q})}$ and $\det{\Delta (\mu_{\rm q})}$ are real 
as shown in \eqref{eq:det_Delta_i}. 
When $\mu_u$ and $\mu_d$ are real, it is possible to 
prove that 
\begin{eqnarray}
\left\{ \det{\Delta (\mu_{\rm I})}\det{\Delta (-\mu_{\rm I})}\right\}^*
=&\det{\Delta (\mu_{\rm I})}\det{\Delta (-\mu_{\rm I})}. 
\label{eq:det_Delta_iso}
\end{eqnarray}
The product of the two determinants is thus real, whereas 
each of the determinants is not. 
LQCD has hence no sign problem for both real and imaginary $\mu_{\rm I}$. 
Actually, LQCD data are available there; see Refs.~\cite{Kogut2,Cea} 
for real $\mu_{\rm I}$ and Refs.~\cite{Cea,D'Elia-iso} for 
imaginary $\mu_{\rm I}$.

The PNJL model explains the LQCD data qualitatively not only 
at real $\mu_{\rm I}$~\cite{Zhang,Mukherjee,Sasaki-T} and 
but also at imaginary $\mu_{\rm I}$~\cite{Sakai3}; 
the eight-quark interaction newly added improves the agreement between 
the PNJL result and the LQCD data~\cite{Sasaki-T}. 
The EPNJL model reproduces the LQCD data quantitatively 
with no free parameter; note that all the parameters in the EPNJL model 
are fixed at vacuum and at imaginary chemical potential. 
As for real $\mu_{\rm I}$, there exists a tricritical point (TCP), i.e. 
a meeting point between the first- and second-order pion-superfluidity 
phase-transition lines. The location of the TCP is predicted by 
the EPNJL model~\cite{Sakai5} as well as 
the PNJL model~\cite{Zhang, Sasaki-T}, 
the chiral perturbation theory~\cite{Son-Stephanov}, 
the strong coupling QCD \cite{Nishida} and so on \cite{Fraga}.

As for real $\mu_{\rm q}$, LQCD data are available at small $\mu_{\rm q}/T$ 
with the Taylor expansion method~\cite{Allton}. 
The EPNJL model reproduces the LQCD data quantitatively with 
no free parameter~\cite{Sakai:2011fa}. 
The EPNJL model predicts~\cite{Sakai5} 
that there is a critical endpoint (CEP)~\cite{AY,Barducci_CEP} 
of the first-order chiral transition in the $\mu_{\rm q}$-$T$ plane.

LQCD has no sign problem also at real isospin and imaginary quark 
chemical potentials, because    
\begin{eqnarray}
&\left\{ \det{\Delta (\mu_{\rm I}+i\Theta T)}\det{\Delta (-\mu_{\rm I}+i\Theta T)}\right\}^*
\nonumber\\
&=\det{\Delta (\mu_{\rm I}+i\Theta T)}\det{\Delta (-\mu_{\rm I}+i\Theta T)}  
\label{eq:det_Delta_iso_Theta}
\end{eqnarray}
for $\mu_u=\mu_{\rm I}+i\Theta T$ and $\mu_d=-\mu_{\rm I}+i\Theta T$. 
The reliability of the PNJL and EPNJL models can be checked 
also in this case, although 
no LQCD data is available at the present stage. 
It should be noted that the LQCD calculation with real $u_{\rm I}$ and imaginary $\mu_{\rm q}$ is equivalent to the LQCD calculation with complex quark number chemical potential under the phase quenched approximation~\cite{NN}.

There is a debate whether the chiral and deconfinement transitions 
at $\mu_{\rm q}=\mu_{\rm I}=0$ coincide or not; 
see Ref.~\cite{Borsanyi} and references therein. 
If the two transitions do not coincide, exotic phases such as 
the constituent-quark phase~\cite{Cleymans,Kouno1} or the 
quarkyonic phase~\cite{McLerran1,Hidaka,Fukushima2,Abuki2,McLerran_largeNc} 
may appear. However, the chiral and deconfinement transitions are 
confirmed to be crossover with LQCD~\cite{KL1994,YAoki_nature}, so that 
transition temperatures of the two crossovers are not well defined. 
The coincidence problem thus includes conceptual difficulty.

As a way of circumventing the conceptual difficulty 
in the coincidence problem, we should consider exact phase transitions 
relevant to the chiral and deconfinement crossovers. 
This is really possible. It is reported with the EPNJL model~\cite{Kouno_CP} 
that the chiral and deconfinement transitions coincide when 
the parity ($P$) restoration and the charge-conjugation ($C$) violation 
occur simultaneously at $\theta=\Theta =\pi$, 
where $\theta$ is the parameter of 
the $\theta$-vacuum~\cite{FIK,Boer,Boer2,Mizher}. 
The chiral and deconfinement crossovers at $\mu_{\rm q}=\mu_{\rm I}=0$ are 
remnants of the $P$ restoration and the $C$ violation at $\theta=\Theta =\pi$, 
respectively. LQCD has, however, the sign problem at finite $\theta$ and 
hence it is difficult to check the validity of the model prediction 
with LQCD. 
Such a problem does not appear at $\mu_{\rm I} >0$ and $\Theta =\pi$, 
as mentioned above.

In this paper, we investigate the interplay 
between the pion-superfluidity and RW phase transitions in 
the region of $\Theta=\pi$ and $\mu_{\rm I} >m_\pi /2$, 
using the PNJL and EPNJL models; here $m_\pi$ 
is the pion mass in the vacuum. 
In the region, the $P$ violation due to the pion-superfluidity transition 
occurs at lower $T$, whereas 
the $C$ violation due to the RW transition takes place at higher $T$. 
These transitions are exact phase transitions and hence 
we can define the order parameters without any ambiguity. 
We also investigate the relation between 
the chiral transition at $\mu_{\rm I}=\Theta=0$ 
and the $P$ violation 
at $\Theta =0$ and $\mu_{\rm I} >m_\pi /2$ by varing $\mu_{\rm I}$ 
and the relation between the deconfinement transition 
at $\mu_{\rm I}=\Theta=0$ and the $C$ violation 
at $\Theta=\pi$ and $\mu_{\rm I} =0$ by varing $\Theta$. 
Finally we discuss how the correlation 
between the $P$ and $C$ violations at $\Theta =\pi$ and $\mu_{\rm I} >m_\pi /2$ is related to 
the correlation between the chiral and deconfinement crossovers 
at $\mu_{\rm I}=\Theta=0$. 
In future, we can check the model prediction made in this paper by using 
LQCD, since LQCD has no sign problem in the region 
and hence LQCD simulations are feasible.

This paper is organized as follows. 
In section II, we explain the PNJL and EPNJL models. 
In section III, numerical results are shown. 
Section IV is devoted to summary.

\section{PNJL model}
\label{PNJL}
The two-flavor PNJL Lagrangian in Euclidean space-time is 
\bea
\mathcal{L}&=&
\bar{q}(\gamma_\nu D_\nu - {\tilde \mu}\gamma_4 + {\tilde m}_0 )q 
-G_{\rm s}\left[
(\bar{q}q)^{2}+(\bar{q}i\gamma_{5} \vec{\tau} q)^{2}
\right]
 \notag \\
&+&{\cal U}(\Phi [A],{\Phi} [A]^*,T), 
\label{eq:E1}
\eea
where $D_\nu=\partial_\nu-iA_\nu$ and 
$A_\nu=\delta_{\nu,4}gA_{4,a}{\lambda^a\over{2}}$
with the gauge field $A_{\nu,a}$, 
the Gell-Mann matrix $\lambda_a$ and the gauge coupling $g$. 
In the NJL sector $G_{\rm s}$ denotes the coupling constant of the scalar-type four-quark interaction. 
As is seen later, the Polyakov potential ${\cal U}$ is a function of the Polyakov loop $\Phi$ and its Hermitian conjugate $\Phi^*$. 

The quark mass matrix ${\tilde m}_0$ is given by ${\tilde m}_0={\rm diag}(m_0, m_0)$. 
The chemical potential matrix ${\tilde \mu}$ is 
defined by ${\tilde \mu}={\rm diag}(\mu_u, \mu_d)$ with 
the $u$-quark ($d$-quark) number chemical potential $\mu_{u}$ ($\mu_{d}$). 
This is equivalent to introducing the baryon and isospin chemical potentials, 
$\mu_{\rm B}$ and $\mu_{\rm iso}$, 
coupled respectively to the baryon charge ${\hat B}$ and to the isospin 
charge ${\hat I_3}$:
\bea
{\tilde \mu}=\mu_{\rm q} \tau_0 + \mu_{\rm I} \tau_3 
\eea
with 
\begin{align}
\mu_{\rm q}=\frac{\mu_{u}+\mu_{d}}{2}=\frac{\mu_{\rm B}}{3},
~~\mu_{\rm I}=\frac{\mu_{u}-\mu_{d}}{2}=\frac{\mu_{\rm iso}}{2} , 
\end{align}
where $\tau_0$ is the unit matrix and $\tau_i$ ($i=1, 2, 3$) 
are the Pauli matrices in the flavor space. 
Note that $\mu_{\rm q}$ is the quark chemical potential and 
$\mu_{\rm I}$ is half the isospin chemical potential ($\mu_{\rm iso}$). 
In the limit of $m_0=\mu_{\rm I}=0$, 
the PNJL Lagrangian has the $SU_{\rm L}(2) \times SU_{\rm R}(2)
\times U_{\rm v}(1) \times SU_{\rm c}(3)$  symmetry. 
For $m_0 \neq 0$ and $\mu_{\rm I} \neq 0$, 
it is reduced to $U_{\rm I_3}(1) \times U_{\rm v}(1) \times SU_{\rm c}(3)$.

The Polyakov loop operator $\hat{\Phi}$ and its Hermitian conjugate 
$\hat{\Phi}^{\dagger}$ are defined as
\begin{eqnarray}
\hat{\Phi}        &=& {1\over{N}} {\rm Tr} L ,~~~~
\hat{\Phi}^{\dagger}  = {1\over{N}} {\rm Tr}L^\dag ,
\end{eqnarray}
with
\begin{eqnarray}
L({\bf x})  &=& {\cal P} \exp\Bigl[
                {i\int^\beta_0 d \tau A_4({\bf x},\tau)}\Bigr],
\end{eqnarray}
where ${\cal P}$ is the path ordering and $A_4 = i A_0 $. 
In the PNJL model, the vacuum expectation values,  
$\Phi=\langle \hat{\Phi} \rangle$ and 
$\Phi^{*}=\langle \hat{\Phi}^{\dagger}  \rangle$, 
are treated as classical variables.  
In the Polyakov gauge, $L$ can be written in a diagonal form 
in color space~\cite{Fukushima}: 
\begin{align}
L 
=  e^{i \beta (\phi_3 \lambda_3 + \phi_8 \lambda_8)}
= {\rm diag} (e^{i \beta \phi_a},e^{i \beta \phi_b},
e^{i \beta \phi_c} ),
\label{eq:E6}
\end{align}
where $\phi_a=\phi_3+\phi_8/\sqrt{3}$, $\phi_b=-\phi_3+\phi_8/\sqrt{3}$
and $\phi_c=-(\phi_a+\phi_b)=-2\phi_8/\sqrt{3}$. 

The Polyakov loop $\Phi$ is an exact order parameter of the spontaneous 
${\mathbb Z}_3$ symmetry breaking in the pure gauge theory.
Although the ${\mathbb Z}_3$ symmetry is not exact 
in the system with dynamical quarks, it still seems to be a good indicator of 
the deconfinement phase transition. 
Therefore, we use $\Phi$ to define the deconfinement phase transition.

The spontaneous breakings of the chiral and the $U_{\rm I_3}(1)$ symmetry are 
described 
by the chiral condensate $\sigma = \langle \bar{q}q \rangle$ and the charged 
pion condensate~\cite{BCPR,He,Zhang,Sasaki-T}
\begin{align}
\pi^{\pm}=\frac{\pi}{\sqrt{2}}e^{\pm i \varphi}
=\langle \bar{q}i \gamma_5 \tau_{\pm}q \rangle ,
\label{charged-pion}
\end{align}
where $\tau_{\pm}=(\tau_1\pm i \tau_2)/\sqrt{2}$.
Since the phase $\varphi$ represents the direction 
of the $U_{\rm I_3}(1)$ symmetry breaking, 
we take $\varphi=0$ for convenience. The pion 
condensate is then expressed by 
\begin{align}
\pi=\langle \bar{q}i \gamma_5 \tau_{1}q \rangle.
\label{pion}
\end{align}
Making the mean field (MF) approximation~\cite{Kashiwa1,Zhang,Sasaki-T}, 
one can obtain the MF Lagrangian as 
\begin{align}
 {\cal L}_{\rm MF}  &=& {\bar q}(\gamma_\nu D_\nu - {\tilde \mu}\gamma_4 + 
              M\tau_0 + N i \gamma_5 \tau_{1})q  ~~~~~~\notag\\
            &\hspace{3mm}&  + G_{\rm s}[\sigma^2 +\pi^2] 
              + {\cal U} \quad
             \label{MF-L}
\end{align}
with 
\begin{eqnarray}
M&=&m_{0} - 2G_{\rm s} \sigma  , 
\label{MM} \\
N&=&- 2G_{\rm s} \pi . 
\label{NN}
\end{eqnarray}
Performing the path integral in the PNJL partition function 
\begin{align}
Z_{\rm PNJL}=\int Dq D\bar{q}
\exp\left[ - \int d^4 x {\cal L}_{\rm MF} \right] , 
\label{PNJL-Z}
\end{align}
we can get the thermodynamic potential $\Omega$ 
(per unit volume), 
\begin{align}
\Omega &=-T\ln(Z_{\rm PNJL})/V
= -2\sum_{i=\pm}\int \frac{d^3{\rm p}}{(2\pi)^3}
         \Bigl[ 3 E_{i}({\rm p}) \nonumber\\
       & + \frac{1}{\beta}
         \ln~ [1 + 3(\Phi+\Phi^{*} e^{-\beta E_{i}^-({\bf p})}) 
        e^{-\beta E_{i}^-({\bf p})}+ e^{-3\beta E_{i}^- ({\bf p})}]
         \nonumber\\
       & + \frac{1}{\beta} 
           \ln~ [1 + 3(\Phi^{*}+{\Phi e^{-\beta E_{i}^+({\bf p})}}) 
            e^{-\beta E_{i}^+({\bf p})}+ e^{-3\beta E_{i}^+({\bf p})}]
           \Bigl]\nonumber\\
       & +G_{\rm s}[\sigma^2 +\pi^2]
         +{\cal U} 
\label{eq:E12-pi} 
\end{align}
with $E_{\pm}^\pm({\rm p})=E_{\pm}({\rm p})\pm \mu_{\rm q}$, 
where  
\bea
E_{\pm}({\rm p})
=\sqrt{(E({\rm p})\pm\mu_{\rm I})^2+N^2} 
\eea
for $E({\rm p})=\sqrt{{\bf p}^2+M^2}$. 
On the right-hand side of \eqref{eq:E12-pi}, only the first term diverges, and 
it is then regularized by the three-dimensional momentum 
cutoff $\Lambda$~\cite{Fukushima,Ratti}. 

We use ${\cal U}$ of Ref.~\cite{Rossner} that is fitted to LQCD data 
in the pure gauge theory at finite $T$~\cite{Boyd,Kaczmarek}: 
\begin{align}
&{\cal U} = T^4 \Bigl[-\frac{a(T)}{2} {\Phi}^*\Phi\notag\\
      &~~~~~+ b(T)\ln(1 - 6{\Phi\Phi^*}  + 4(\Phi^3+{\Phi^*}^3)
            - 3(\Phi\Phi^*)^2 )\Bigr], \label{eq:E13}\\
&a(T)   = a_0 + a_1\Bigl(\frac{T_0}{T}\Bigr)
                 + a_2\Bigl(\frac{T_0}{T}\Bigr)^2,
 ~~~b(T)=b_3\Bigl(\frac{T_0}{T}\Bigr)^3 , \label{eq:E14}
\end{align}
where parameters are summarized in Table \ref{table-para}.  
The Polyakov potential yields a first-order deconfinement phase transition 
at $T=T_0$ in the pure gauge theory.
The original value of $T_0$ is $270$ MeV determined from the pure gauge 
LQCD data, but the PNJL model with this value of $T_0$ yields a larger 
value of the pseudocritical temperature $T_\mathrm{c}$ 
at zero chemical potential than $T_{\rm c}=173 \pm 8$~MeV predicted by 
the full LQCD simulation~\cite{KL1994,Karsch4,Kaczmarek2}. 
We then reset $T_0$ to 212~MeV~\cite{Sakai2} so as to 
reproduce the LQCD result. 

\begin{table}[h]
\begin{center}
\begin{tabular}{llllll}
\hline \hline
~~~~~$a_0$~~~~~&~~~~~$a_1$~~~~~&~~~~~$a_2$~~~~~&~~~~~$b_3$~~~~~
\\
\hline
~~~~$3.51$ &~~~~$-2.47$ &~~~~$15.2$ &~~~~$-1.75$\\
\hline \hline
\end{tabular}
\caption{
Summary of the parameter set in the Polyakov-potential sector 
determined in Ref.~\cite{Rossner}. 
All parameters are dimensionless. 
}
\label{table-para}
\end{center}
\end{table}

Table \ref{NJL-para} shows parameters in the NJL sector used in 
the present analyses. 
This set can reproduce the pion decay constant $f_{\pi}=93.3$~MeV and the pion mass $m_{\pi}=138$~MeV at vacuum ($T=\mu_{\rm q}=\mu_{\rm I}=0$). 

\begin{table}[h]
\begin{center}
\begin{tabular}{lcccc}
\hline \hline
~~&$G_s$ & $m_0$ & $\Lambda$
\\
\hline
~~&~$5.498~[{\rm GeV}^{-2}]$~&~$5.5~[{\rm MeV}]$~ & ~$631.5~[{\rm MeV}]$~ \\
\hline \hline
\end{tabular}
\caption{
Summary of parameters in the NJL sector. 
}
\label{NJL-para}
\end{center}
\end{table}

The classical variables $X=\Phi$, ${\Phi}^*$, $\sigma$ and $\pi$ 
are determined by the stationary conditions 
\begin{align}
\partial \Omega/\partial X=0. 
\label{eq:SC}
\end{align}
The solutions to the stationary conditions do not give 
the global minimum of $\Omega$ 
necessarily. There is a possibility 
that they yield a local minimum or even 
a maximum. We then have checked that the solutions yield 
the global minimum when  
the solutions 
$X(T,\mu_{\rm q},\mu_{\rm I})$
are inserted into (\ref{eq:E12-pi}). 

The thermodynamic potential (\ref{eq:E12-pi}) is invariant under the parity transformation 
\begin{eqnarray}
\pi \to -\pi. 
\label{parity}
\end{eqnarray}
Parity odd quantity such as the pion condensate is an order parameter 
of $P$ symmetry. When pion condensates, $P$ symmetry is spontaneously 
broken~\cite{Son-Stephanov}. 

When $\mu_{\rm I}\neq 0$, charge-conjugation symmetry
is explicitly broken even at $\mu_{\rm q}=0$~\cite{Kouno,Kouno_CP}, 
since $u$ and $d$ quarks have different electric charge. 
In the present system where the electromagnetic field is switched off, however, we can neglect the electric charges and consider only 
a charge-conjugation transformation associated with baryon and color charges. 
We refer to symmetry under the transformation as 
partial charge-conjugation symmetry ($\tilde{C}$) in this paper. 
As far as $\Omega$ of (\ref{eq:E12-pi}) is concerned, 
the transformation is simply represented by 
\begin{eqnarray}
\phi \to -\phi 
\label{charge}
\end{eqnarray}
with $\phi$ the phase of the Polyakov loop $\Phi$. 
When $\mu_{\rm q}=i \Theta T=i n \pi T$ for any integer $n$, 
$\Omega$ is $\tilde{C}$-invariant. As shown below, $\tilde{C}$ symmetry is 
spontaneously broken at higher temperature for odd $n$. 

The thermodynamic potential $\Omega$ of \eqref{eq:E12-pi} 
has a trivial periodicity of $2\pi$ in $\Theta$. 
We then mainly consider one circle, $0 \le \Theta \le 2\pi$. 
In addition, $\Omega$ has the RW periodicity~\cite{RW}. 
The thermodynamical potential $\Omega$ is not 
invariant under the ${\mathbb Z}_3$ transformation, 
\begin{eqnarray}
\Phi  \to \Phi e^{-i{2\pi k/{3}}} \;,\quad
\Phi ^{*} \to \Phi^{*} e^{i{2\pi k/{3}}} \;, 
\end{eqnarray}
for any integer $k$, 
while ${\cal U}$ of (\ref{eq:E13}) is invariant. 
Instead of the ${\mathbb Z}_3$ symmetry, however, $\Omega$ is invariant 
under the extended ${\mathbb Z}_3$ transformation~\cite{Sakai}, 
\begin{align}
&e^{\pm i \Theta} \to e^{\pm i \left( \Theta +{2\pi k\over{3}}\right)},\quad  
\Phi \to \Phi e^{-i{2\pi k\over{3}}}, 
\notag\\
&\Phi^{*} \to \Phi^{*} e^{i{2\pi k\over{3}}} .
\label{eq:K2}
\end{align}
The extended ${\mathbb Z}_3$ symmetry guarantees the RW periodicity, 
as shown below. 
The thermodynamic potential (\ref{eq:E12-pi}) can be 
rewritten with the modified Polyakov loop
\begin{eqnarray}
\Psi =\Phi e^{i\Theta}
\label{Psi} 
\end{eqnarray}
invariant under the extended-${\mathbb Z}_3$ transformation: 
\begin{align}
\Omega &=-T\ln(Z_{\rm PNJL})/V
= -2\sum_{i=\pm}\int \frac{d^3{\rm p}}{(2\pi)^3}
         \Bigl[ 3 E_{i}({\rm p}) \nonumber\\
       & + \frac{1}{\beta}
         \ln~ [1 + 3\Psi e^{-\beta {E_i}({\bf p})} \nonumber\\
       &
       +3\Psi^{*} e^{-2\beta E_{i}({\bf p})}e^{i3\Theta} 
        + e^{-3\beta E_{i}({\bf p})}e^{3i\Theta}]
         \nonumber\\
       & + \frac{1}{\beta}
         \ln~ [1 + 3\Psi^{*} e^{-\beta {E_i}({\bf p})} \nonumber\\
       &
       +3\Psi e^{-2\beta E_{i}({\bf p})}e^{-i3\Theta} 
        + e^{-3\beta E_{i}({\bf p})}e^{-3i\Theta}]
        +G_{\rm s}[\sigma^2 +\pi^2]
\nonumber\\
       &+T^4 \Bigl[-\frac{a(T)}{2} {\Psi}^*\Psi\notag\\
       &+ b(T)\ln(1 - 6{\Psi\Psi^*}  + 4(\Psi^3e^{-3i\Theta}+{\Psi^*}^3e^{3i\Theta})
            - 3(\Psi\Psi^*)^2 )\Bigr]. 
\label{Omega}
\end{align}
Obviously, $\Omega$ is extended-${\mathbb Z}_3$ invariant, 
since it depends on $\Theta$ only through the factor $e^{3i\Theta}$. 
The factor also guarantees that 
\begin{eqnarray}
\Omega(T,\Theta )=\Omega (T,\Theta +{2\pi\over{3}})=\Omega (T,\Theta +{4\pi\over{3}}). 
\label{RW-periodicity}
\end{eqnarray}
Thus, $\Omega (\Theta)$ has a shorter period ${2\pi/3}$ in $\Theta$. 
This periodicity was first found by Roberge and Weiss~\cite{RW} 
for QCD and now it is called 
the Roberge-Weiss (RW) periodicity. 
Using the perturbation theory and the strong-coupling lattice theory, 
Roberge and Weiss also found that a first-order phase transition 
occurs at $\Theta =\pi/3$ mod $2\pi/3$ when $T$ is higher than 
some critical value $T_{\rm C}$. 
The transition is also called the Roberge-Weiss transition. 
After the theoretical prediction, the RW periodicity and the RW transition 
were confirmed by LQCD~\cite{FP,Elia,Chen,D'Elia-iso,Cea,D'Elia-3,FP2010,Nagata,Takaishi}. 
Recently, the RW periodicity and the RW transition were also confirmed by 
Holographic QCD~\cite{Aarts}. 
The RW transition induces 
the $C$ breaking~\cite{Kouno,Kouno_CP} for $\mu_{\rm I}=0$ and 
the $\tilde{C}$ breaking for $\mu_{\rm I}\neq 0$. 
The $C$ and $\tilde{C}$ breakings occur at $\Theta=\pi$ and 
the ${\mathbb Z}_3$ images of the breakings also appear 
at $\Theta=i{\pi/3}$ and $i{5\pi/3}$. 
It is convenient to use the phase $\psi$ 
of the modified Polyakov loop $\Psi$ as an order parameter 
of the $C$ and $\tilde{C}$ breakings~\cite{Kouno,Kouno_CP}. 
When $C$ or $\tilde{C}$ symmetry is preserved, $\psi$ vanishes; 
here we consider $\psi$ in the region $-\pi \le \psi \le \pi$.

In LQCD simulations at imaginary $\mu_{\rm q}$, 
the chiral and deconfinement transitions 
take place simultaneously, 
although they are crossover~\cite{FP,Elia,Chen,D'Elia-iso,Cea,D'Elia-3,FP2010,Nagata,Takaishi}. 
Such a strong correlation can not be reproduced by 
the PNJL model\cite{Sakai,Sakai2,Sakai3}. 
This problem is solved by the entanglement PNJL (EPNJL) model~\cite{Sakai5}. 
In the EPNJL model, 
the four-quark vertex depends on the Polykov loop. 
In QCD, the four-quark vertex is generated by the gluon propagation 
between two quarks. If the gluon has a finite vacuum expectation value 
in its time component, the gluon propagation can depend on $\Phi$ 
through the vacuum expectation value~\cite{Sakai5}. 
In fact, recent calculations~\cite{Braun,Kondo,Herbst} of 
the exact renormalization group equation (ERGE)~\cite{Wetterich} suggest 
that mixing interactions between $\sigma$ and $\Phi$ are induced. 
It is highly expected that the functional form and the strength of 
the entanglement vertex are determined in future by ERGE. 
In Ref.~\cite{Sakai5}, the following $\Phi$ dependence of $G$ is assumed 
by respecting the chiral symmetry, 
$P$ symmetry, $C$ symmetry and the extended $\mathbb{Z}_3$ symmetry: 
\begin{eqnarray}
G(\Phi)=G[1-\alpha_1\Phi\Phi^*-\alpha_2(\Phi^3+\Phi^{*3})]. 
\label{entanglement-vertex}
\end{eqnarray}
In the EPNJL model, the chiral and deconfinement crossovers coincide 
at zero chemical potential. 
The EPNJL model with the parameter set, 
$\alpha_1=\alpha_2=0.2$ and $T_0=190$~MeV, can reproduce 
LQCD data at imaginary chemical potential~\cite{Sakai5}, 
real isospin chemical potential~\cite{Sakai5} and 
small $\mu_{\rm q}/T$~\cite{Sakai:2011fa}. 
The agreement between LQCD data  and the EPNJL result 
persists also under strong magnetic field~\cite{D'Elia5,Gatto:2010pt}. 
We then use the EPNJL model with this parameter set also in this paper. 

The entanglement vertex $G(\Phi )$ agrees with 
the parameter $G$ at $T=0$ since $\Phi=0$ there. 
At vacuum the EPNJL model is then reduced to the NJL model; 
note that the PNJL model agrees with the NJL model at vacuum. 
As a merit of this property, the NJL sector of the EPNJL model has 
the same values of parameters as the NJL model. 

Very recently, it was shown that the coincidence problem 
between the chiral and deconfinement transitions does not occur 
also in the nonlocal PNJL model~\cite{Kashiwa_NL_IM}. In the model, 
the entanglement between $\sigma$ and $\Phi$ naturally appears 
in the renormalization factor. It is then interesting as a future work 
to study the relation between the nonlocal PNJL and the EPNJL model.

\section{Numerical Results}
\label{NumericalResults}
\subsection{Results at $\mu_{\rm I}=0$ and $\mu_{q}=0$}
\label{mui000theta000_SubS}

\begin{figure}[htbp]
\begin{center}
 \includegraphics[width=0.5\textwidth]{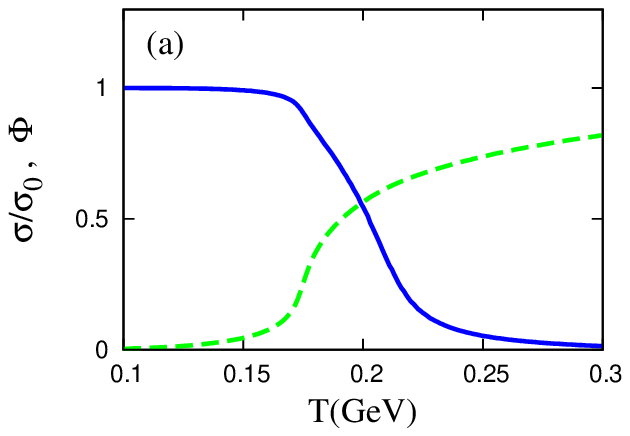}
 \includegraphics[width=0.5\textwidth]{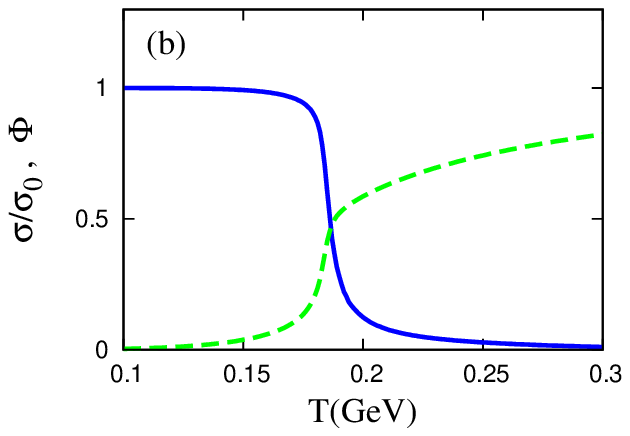} 
\end{center}
\caption{(color online). $T$ dependence of the chiral condensate $\sigma$ 
and the Polyakov loop $\Phi$ at $\mu_{\rm I}=0$ and $\mu_{\rm q}=0$ 
calculated with (a) the PNJL model and (b) the EPNJL model. 
The solid (dashed) line represents $\sigma$ ($\Phi$). 
Here, $\sigma$ is normalized by the value $\sigma_0$ at vacuum. 
 }
\label{mui000theta000}
\end{figure}

In this subsection, we consider the case of $\mu_{\rm I}=\mu_{\rm q}=0$ 
and $m_0=5.5$~MeV. The chiral condensate $\sigma$ and 
the Polykov loop $\Phi$ are approximate order parameters 
of the chiral and deconfinement transitions, respectively. 
Figure\ref{mui000theta000}(a)  shows $T$ dependence of $\sigma$ and $\Phi$ 
calculated with the PNJL model.  
Both the chiral restoration and the deconfinement transition are seen to be 
crossover. 
The chiral restoration starts at the same temperature 
($T\approx 170$~MeV) as the deconfinement transition 
does, but the former transition is slower than the latter transition.  
Consequently, the pseudocritical temperature $T_{\sigma}$ 
of the chiral restoration 
becomes higher than that $T_{\Phi }$ of the deconfiment transition. 
Actually, it is found that 
$T_{\sigma}=216$~MeV and $T_{\Phi }=173$~MeV~\cite{Sakai2}, 
when the pseudocritical temperatures are defined by 
the peak positions of $\sigma$ and $\Phi$, respectively~\cite{Kashiwa1}. 
Figure\ref{mui000theta000}(b)  shows $T$ dependence of $\sigma$ and $\Phi$ 
calculated with the EPNJL model. 
The chiral restoration occurs with the same speed as the deconfinement 
transition, so that $T_{\sigma}=T_{\Phi }=173$~MeV~\cite{Sakai5}. 

When two transitions are first order, they might coincide exactly~\cite{BCGG}. 
At zero $\mu_{\rm q}$, however, the chiral and deconfinement transitions 
are found to be crossover \cite{KL1994,YAoki_nature}. 
There is hence no a priori reason why the two crossovers coincide exactly. 
In LQCD simulations at zero chemical 
potential~\cite{KL1994,SAoki1998,YAoki2006}, actually, 
there is a debate whether the transitions really coincide or not; 
see Ref.~\cite{Borsanyi} and references therein.

The coincidence problem has conceptual difficulty, since 
$\sigma$ and $\Phi$ are approximate order parameters 
of the chiral and deconfinement transitions. 
This point can be circumvented by considering exact phase transitions 
relevant to the chiral and deconfinement crossovers. 
In the previous work~\cite{Kouno_CP}, we found 
that the chiral and deconfinement transitions 
coincide when the $P$ restoration 
and the $C$ violation occur simultaneously 
at $\theta =\Theta=\pi$. 
The chiral restoration at $\theta =\Theta=0$ is a remnant of 
the $P$ restoration at $\theta =\Theta=\pi$, whereas 
the deconfinement crossover at $\theta =\Theta=0$ is a remnant of 
the $C$ violation at $\theta =\Theta=\pi$. 
Note that the $P$ restoration and $C$ violation are exact phase transitions. 
This theoretical prediction, however, cannot be confirmed by LQCD, 
since LQCD has the sign problem at finite $\theta$. 
The region of $\mu_{\rm I}>m_\pi/2$ and $\Theta =\pi$ that we consider 
in this paper does not suffer from the sign problem.

\subsection{Results at finite $\mu_{\rm I}$ and $\mu_{q}=0$}
\label{mui138theta000_SubS}

\begin{figure}[htbp]
\begin{center}
  \includegraphics[width=0.5\textwidth]{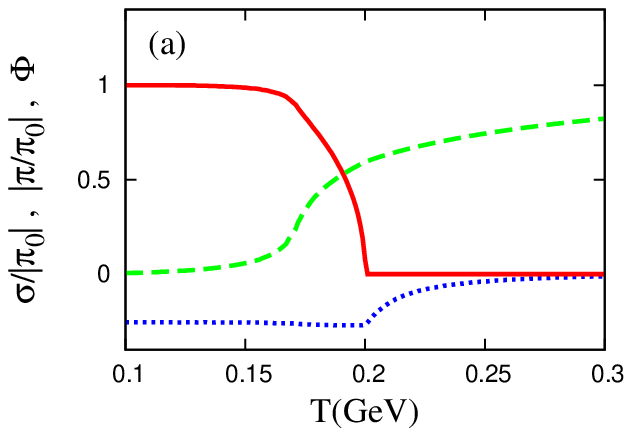}
  \includegraphics[width=0.5\textwidth]{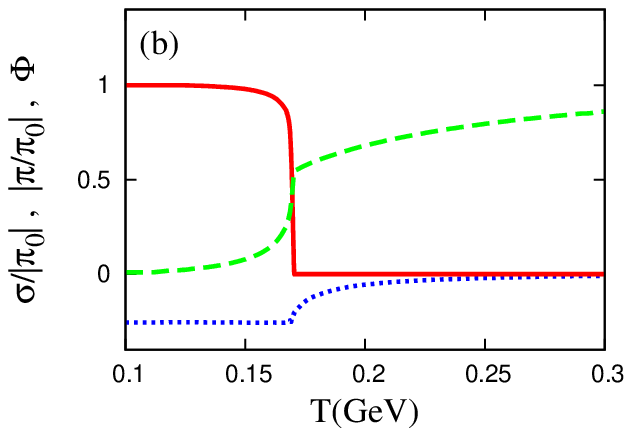}
\end{center}
\caption{(color online). 
$T$ dependence of chiral condensate $\sigma$, pion condensation $\pi$ and Polyakov loop $\Phi$ at $\mu_{\rm I}=m_\pi$ and $\mu_{\rm q} =0$ 
calculated with (a) the PNJL model and (b) the EPNJL model. 
The solid (dashed, dotted) line represents $\pi$ ($\Phi$, $\sigma$). 
Here $\sigma$ and $\pi$ are normalized by $|\pi_0|$, where $\pi_0=\pi(T=0)$. 
 }
\label{mui138theta000}
\end{figure}

When $\mu_{\rm I}>m_\pi/2$, the pion condensation becomes finite 
at low temperature, so that $P$ symmetry is spontaneously broken 
there~\cite{Son-Stephanov}. In contrast, 
the pion condensate $\pi$ is zero at $\mu_{\rm I} \le m_\pi /2$, so that 
$P$ symmetry is preserved there. 
Figure \ref{mui138theta000} shows $T$ dependence of the pion condensate 
at $\mu_{\rm I}=m_\pi$ and $\mu_q=0$. 
The pion condensation disappears at $T > T_P=201(170)$~MeV 
in the PNJL (EPNJL) model.

\begin{figure}[htbp]
\begin{center}
 \includegraphics[width=0.5\textwidth]{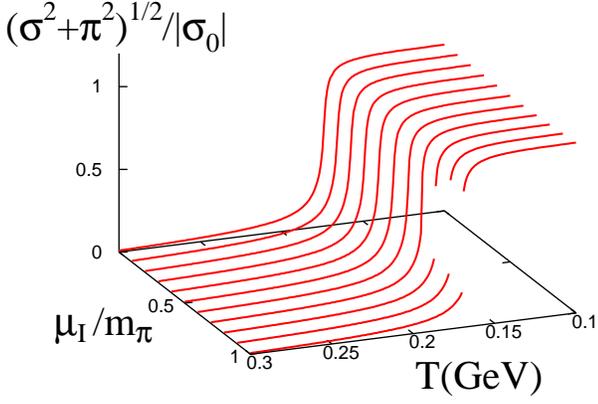}
\end{center}
\caption{(color online). The order parameter $\sqrt{\sigma^2+\pi^2}$ 
of the chiral transition as a function of 
$T$ and $\mu_{\rm I}$ plane calculated with the EPNJL model. 
Here we consider the case of $\mu_{\rm q} =0$. }
\label{T-mui-r}
\end{figure}

Figure \ref{T-mui-r} shows the approximate 
order parameter $\sqrt{\sigma^2+\pi^2}$ of the chiral transition 
as a function of 
$T$ and $\mu_{\rm I}$ calculated with the EPNJL model, where $\Theta=0$. 
As for $\mu_{\rm I}>m_\pi /2$ where $P$ symmetry is broken at small $T$ and 
preserved at large $T$, $\sqrt{\sigma^2+\pi^2}$ has either a cusp 
or a discontinuity, as expected. 
These singularities in $\sqrt{\sigma^2+\pi^2}$ come from the fact that 
the pion condensate $\pi$ has either a cusp or a discontinuity there; 
more precisely, the $P$ restoration is second order 
at $m_\pi /2<\mu_{\rm I}<95$~MeV and first order at $\mu_{\rm I}>95$~MeV. 
As for $\mu_{\rm I}<m_\pi/2$ where the pion condensation 
vanishes and hence $P$ symmetry is preserved, in contrast, 
$\sqrt{\sigma^2+\pi^2}$ changes smoothly. 
However, the change is still rapid, since $\sigma$ has a rapid change. 
The rapid change of $\sigma$ is nothing but the chiral restoration 
at zero and small $\mu_{\rm I}$. 
Thus, the chiral restoration at zero $\mu_{\rm I}$ can be 
regarded as a remnant of the $P$ restoration at $\mu_{\rm I}>m_\pi /2$.

\subsection{Results at $\mu_{\rm I}=0$ and $\Theta =\pi$}
\label{mui000theta100_SubS}

\begin{figure}[htbp]
\begin{center}
 \includegraphics[width=0.5\textwidth]{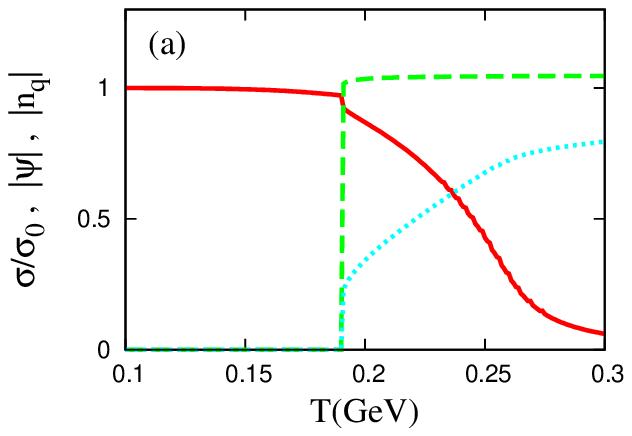}
 \includegraphics[width=0.5\textwidth]{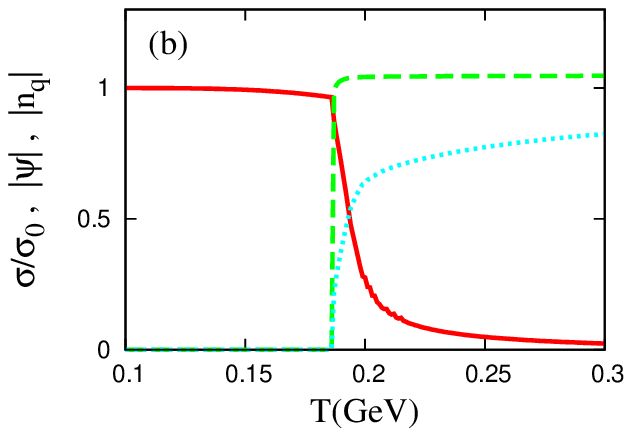}
\end{center}
\caption{(color online). $T$ dependence of $\sigma$, $|\psi|$ and $|n_{\rm q}|$ at $\mu_{\rm I}=0$ and $\Theta =\pi$ calculated with (a) the PNJL model 
and (b) the EPNJL model. 
The solid (dashed, dotted) line represents $\sigma$ ($|\psi|$, 
$|n_q|$). 
Here, $\sigma$ is normalized by the value $\sigma_0$ at vacuum and $n_q$ 
is by $N_f T^3$, where $N_f=2$. 
 }
\label{mui000theta100}
\end{figure}

When $\Theta =\pi $, the RW transition take places at higher $T$. 
Hence $C$ symmetry is spontaneously broken there~\cite{Kouno,Kouno_CP}. 
As an order parameter of the $C$ violation, we can consider 
$\Theta$-odd quantities such as the quark number density $n_{\rm q}$ 
or the phase $\psi$ of the modified Polyakov loop. 
Figure \ref{mui000theta100} shows $T$ dependence 
of $\sigma$, $|\psi|$ and $|n_{\rm q}|$ at $\mu_{\rm I}=0$ and $\Theta =\pi$. 
Note that $n_{\rm q}$ is pure imaginary and hence 
$|n_{\rm q}|=|{\rm Im}(n_{\rm q})|$. 
In the PNJL (EPNJL) model, the $C$ violation occurs at $T>T_C=189 (185) $~MeV, 
since $n_{\rm q}$ and $\psi$ are finite there.

\begin{figure}[htbp]
\begin{center}
 \includegraphics[width=0.5\textwidth]{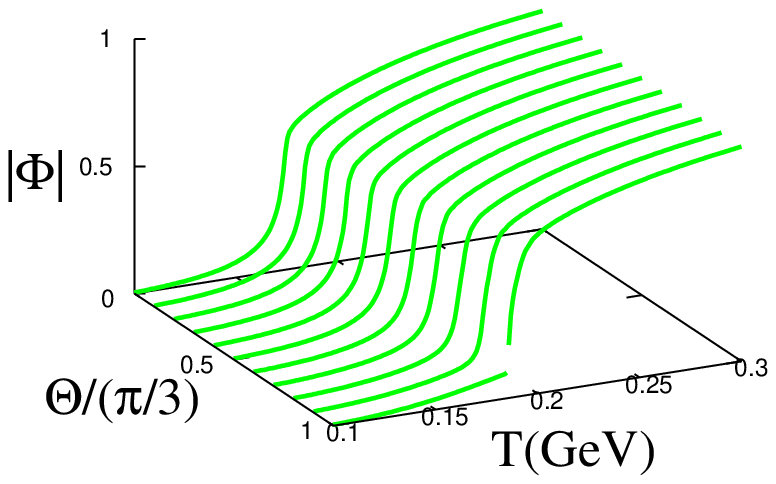}
\end{center}
\caption{(color online). The approximate order parameter $|\Phi|$ 
of the deconfinement transition as a function of $T$ and $\theta$ calculated with 
the EPNJL model. Here the case of $\mu_{\rm I}=0$ is taken. 
}
\label{T-theta-Phi}
\end{figure}

Figure \ref{T-theta-Phi} shows $T$ and $\Theta$ dependences of 
the approximate order parameter $|\Phi|$ 
of the deconfinement transition calculated with the EPNJL model; 
here the case of $\mu_{\rm I}=0$ is considered. 
Since $|\Phi|$ is $\Theta$-even and has the RW periodicity, 
we can consider the region $0 \le \Theta \le \pi/3$ only. 
There appears a first order phase transition at $\Theta =\pi/3$. 
This is the RW phase transition in which $C$ symmetry is spontaneously 
broken. The transition becomes crossover as $\Theta$ decreases from $\pi/3$. 
However, a smooth but rapid change remains even at $\Theta =0$. 
The rapid change is nothing but the deconfinement transition 
at zero $\Theta$. Thus, the deconfinement transition at zero $\mu_{q}$ 
is a remnant of the first-order RW phase transition 
at $\Theta =\pi /3$~\cite{Kouno,Kouno_CP}. 

\subsection{Results at finite $\mu_{\rm I}$ and $\Theta$}
\label{mui138theta100_SubS}

\begin{figure}[htbp]
\begin{center}
 \includegraphics[width=0.5\textwidth]{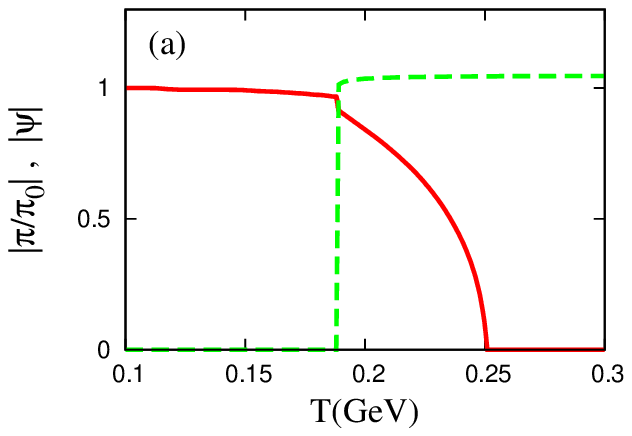}
 \includegraphics[width=0.5\textwidth]{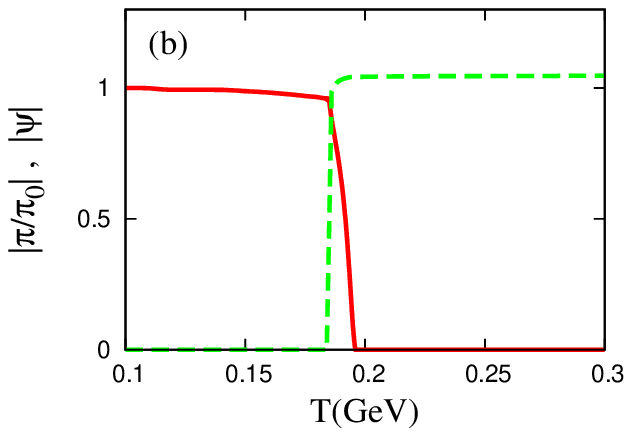}
\end{center}
\caption{(color online). $T$ dependence of the pion condensate $\pi$ and 
the phase $\psi$ of $\Psi$ at $\mu_{\rm I}=m_\pi$ and $\Theta =\pi$ 
calculated with (a) the PNJL model and (b) the EPNJL model. 
The solid (dashed) line represents $|\pi|$ ($|\psi|$) and 
$\pi$ is normalized by the value $\pi_0$ at $T=0$. 
 }
\label{mui138theta100}
\end{figure}

Figure~\ref{mui138theta100} shows $T$ dependence of the pion condensate 
$\pi$ and the phase $\psi$ of $\Psi$
at $\mu_{\rm I}=m_\pi$ and $\Theta =\pi$. 
In the PNJL (EPNL) model, 
$P$ symmetry is spontaneously broken below $T_P=250(196)$MeV 
as shown by a finite value of the pion condensate. In contrast, 
$\tilde{C}$ symmetry is spontaneously violated 
above $T_{\tilde{C}}=188(184)$MeV as indicated by a finite value of $\psi$.  
Thus $P$ and/or $\tilde{C}$ symmetries are always broken at any temperature. 
This situation is the same as that at $\theta =\Theta =\pi$~\cite{Kouno_CP}. 
The difference $\Delta T\equiv T_P-T_{\tilde{C}}$ is smaller 
in the EPNJL model than in the PNJL model. 
The entanglement interaction thus makes $\Delta T$ very small, 
although the two transitions do not coincide exactly. 
Note that at $\mu_{\rm I}=m_\pi$ the $P$ restoration is second order whereas 
the $\tilde{C}$ violation is first order in both the PNJL and EPNJL models. 
Comparing Fig.~\ref{mui138theta100} with Fig.~\ref{mui138theta000}, one can 
see that $T_P$ goes up as $\Theta$ increases. 

\begin{figure}[htbp]
\begin{center}
 \includegraphics[width=0.5\textwidth]{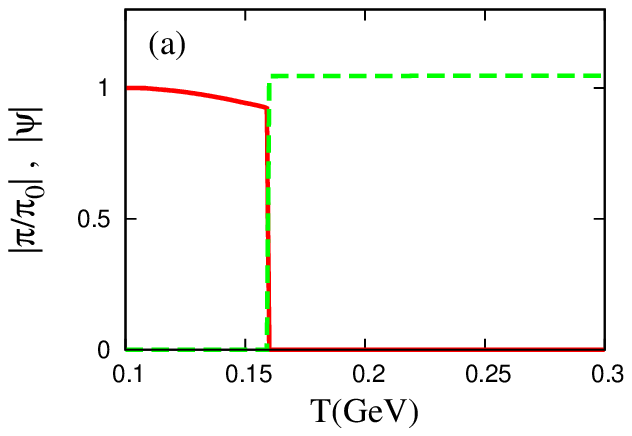}
 \includegraphics[width=0.5\textwidth]{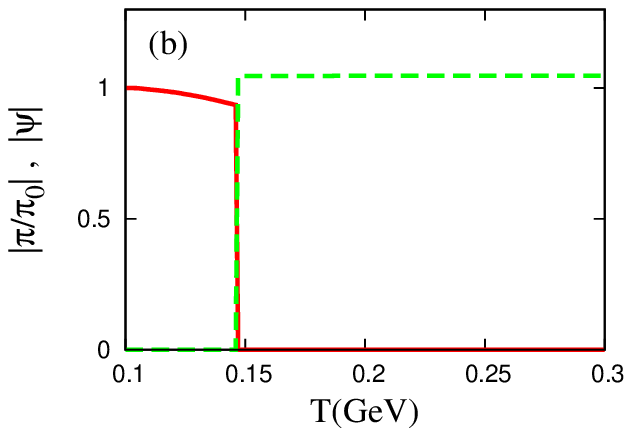}
\end{center}
\caption{(color online). 
$T$ dependence of the pion condensate $\pi$ and 
the phase $\psi$ of $\Psi$ at $\mu_{\rm I}=600$~MeV and $\Theta =\pi$ 
calculated with (a) the PNJL model and (b) the EPNJL model. 
See Fig.~\ref{mui138theta100} for the definition of lines. 
 }
\label{mui600theta100}
\end{figure}

Figure~\ref{mui600theta100} shows the same quantities 
as Fig.~\ref{mui138theta100}, 
except the case of $\mu_{\rm I}=600$MeV is considered. 
In both the PNJL and EPNL models, the $P$ restoration and 
the $\tilde{C}$ violation occur at the same temperature, more precisely 
$T_P=T_{\tilde{C}}=160$~MeV for the PNJL model and 147~MeV for the EPNJL model. This is because the $P$ restoration becomes sharper and 
eventually first-order as $\mu_{\rm I}$ increases. 
Thus the first-order nature of the $P$ restoration makes 
the $\tilde{C}$ violation strong first-order and consequently attracts 
the $\tilde{C}$ violation. 

\begin{figure}[htbp]
\begin{center}
 \includegraphics[width=0.5\textwidth]{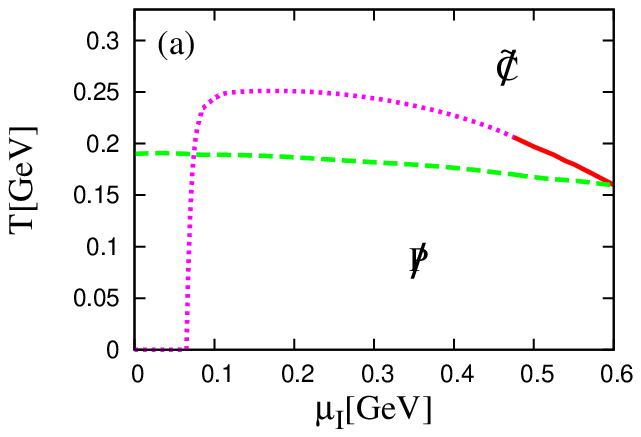}
 \includegraphics[width=0.5\textwidth]{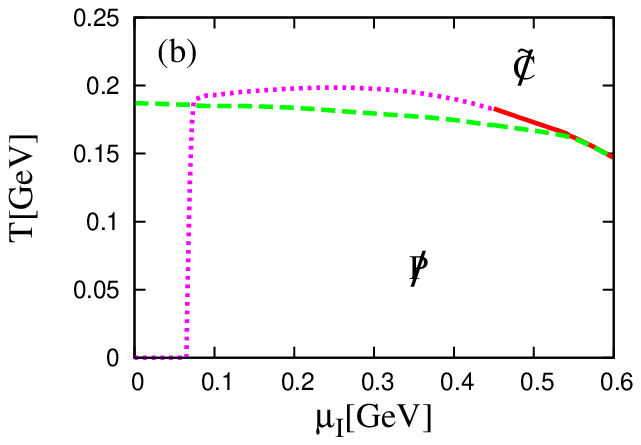}
\end{center}
\caption{(color online). 
Phase diagram in the $\mu_{\rm I}$-$T$ plane at $\Theta =\pi$ 
calculated with 
with (a) the PNJL model and (b) the EPNJL model. 
The dashed (solid) line represents the second (first) order $P$ transition, 
while the dotted line represents the first order $\tilde{C}$ transition. 
 }
\label{muI-T-phase}
\end{figure}

Figure~\ref{muI-T-phase} shows the phase diagram in the $\mu_{\rm I}$-$T$ 
plane at $\Theta =\pi$. 
The $P$ violation due to the pion condensation occur at $\mu_{\rm I}>m_\pi/2$ 
and lower $T$. 
The $P$ restoration with respect to increasing $T$ is 
second-order at $\mu_{\rm I} < 480 (450)$~MeV and first-order 
at $\mu_{\rm I} > 480 (450)$~MeV in the PNJL (EPNJL) model. 
The $\tilde{C}$ violation is always first order. 
When $\mu_{\rm I} > 600(545)$~MeV, the $P$ restoration coincides with 
the $\tilde{C}$ violation in the PNJL(EPNJL) model, as already mentioned 
in Fig.~\ref{mui600theta100}. 
The region where $P$ and $\tilde{C}$ symmetries are violated 
simultaneously is tiny in the EPNJL model 
compared with the case of the PNJL model. 

\begin{figure}[htbp]
\begin{center}
 \includegraphics[width=0.5\textwidth]{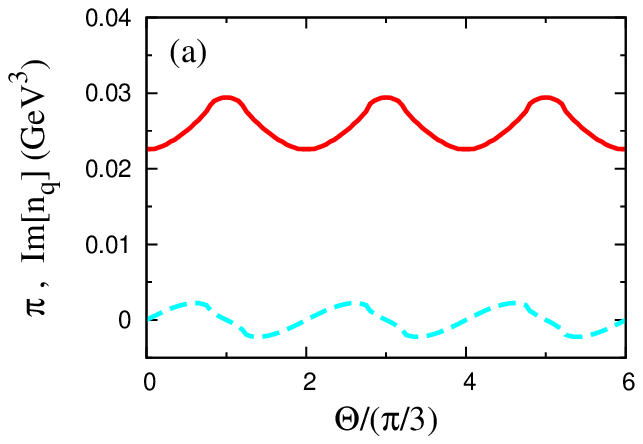}
 \includegraphics[width=0.5\textwidth]{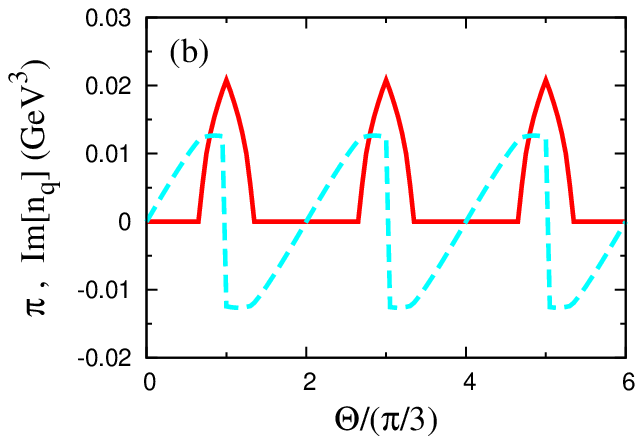}
\end{center} 
\caption{(color online). $\Theta$ dependence of the pion condensate $\pi$ 
 and the imaginary part ${\rm Im}(n_q)$ of quark number density 
 calculated with the PNJL model. Here the case of $\mu_{\rm I}=m_\pi$ is 
 taken. Quantities are shown in GeV$^3$. 
 Panel (a) is for the case of $T=180$~MeV and panel (b) is for the case of 
 $T=220$~MeV.
 The solid (dotted) line represents $\pi$~(${\rm Im}(n_q)$). }
\label{Theta_dep}
\end{figure}

Figure~\ref{Theta_dep} shows $\Theta$ dependence 
of the pion condensation $\pi$ and the imaginary part 
of the quark number density $n_{\rm q}$ 
calculated with the PNJL model; note that 
$n_{\rm q}$ is pure imaginary in this case. 
In panel (a) where $T=180~[{\rm MeV}] < T_{\rm C} =188~[{\rm MeV}]$, 
both $\pi$ and $n_{\rm q}$ are smooth functions of $\Theta$ and 
no RW phase transition is seen at $\Theta=\pi /3$ mod $2\pi /3$. 
In panel (b) where $T=220~[{\rm MeV}] > T_{\rm C}$, 
the $\Theta$-even quantity $\pi$ has a cusp at $\Theta=\pi /3$ mod $2\pi /3$, 
whereas the $\Theta$-odd quantity $n_{\rm q}$ is discontinuous there
~\cite{Kashiwa5}. 
This singularity is called the RW phase transition. 
The pion condensate is finite only around $\Theta=\pi /3$ mod $2\pi /3$ 
and vanishes when $T_P < 220$~MeV. 
This behavior of $\pi$ resembles that of $\sigma$ in the case of $m_0=\mu_{\rm I}=0$~\cite{Sakai_R} where $\sigma$ and $\pi$ are completely symmetric. 

\begin{figure}[htbp]
\begin{center}
 \includegraphics[width=0.5\textwidth]{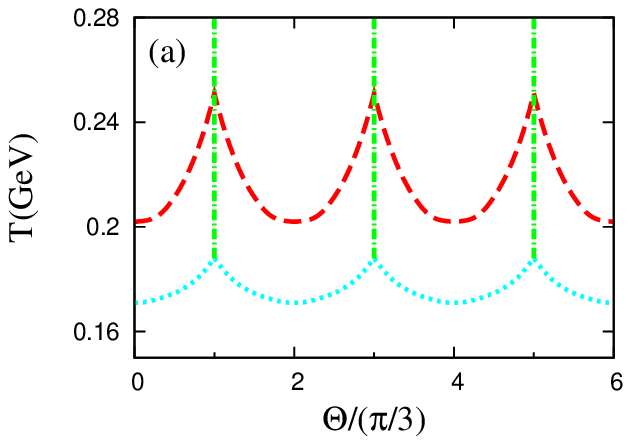}
 \includegraphics[width=0.5\textwidth]{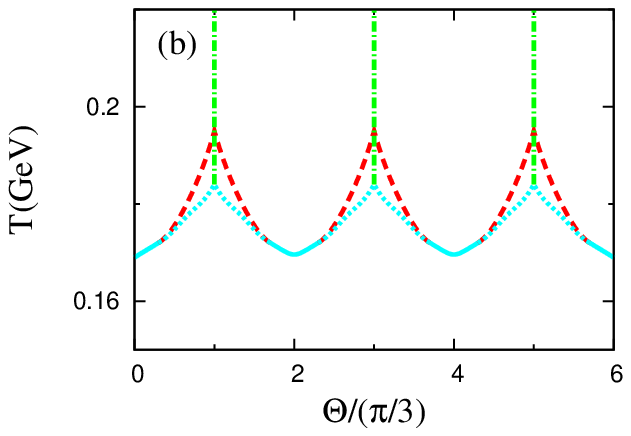}
\end{center}
\caption{(color online). 
Phase diagram in the $\theta$-$T$ plane at $\mu_{\rm I}=m_\pi$ 
calculated with (a) the PNJL model and (b) the EPNJL model.  
The dot-dashed lines stand for the RW  transition, and 
the dotted lines represent the deconfinement transition. 
The solid (dashed) lines denote the $P$ restoration of first-order 
(second-order). 
 }
\label{Theta-T-phase}
\end{figure}

Figure~\ref{Theta-T-phase} shows the phase diagram 
in the $\Theta$-$T$ plane at $\mu_{\rm I}=m_\pi$ calculated with 
(a) the PNJL model and (b) the EPNJL model. 
The phase diagram is $\Theta$-even and has the RW periodicity. 
The RW transition at $\Theta=\pi /3$ mod $2\pi /3$ is first order. 
The deconfinement transition is first-order in the vicinity of 
the endpoint of the RW transition, crossover for other $\Theta$. 
The $P$ restoration is always second order in the PNJL model. 
In the EPNJL model, the $P$ restoration is first-order at 
$\Theta/(\pi /3)=-0.3 \sim 0.3,~1.7\sim 2.3,~3.7\sim 4.3$ and 
second order for other $\Theta$. 
Thus the imaginary quark chemical potential makes 
the $P$ restoration weaker. 
In the regions where the $P$ restoration is first-order, the $P$ restoration and the deconfinement transition almost coincide with each other in the EPNJL model. 

\section{Summary}
\label{Summary}

We have studied the interplay 
between the pion-superfluidity and RW phase transitions
at real isospin and imaginary quark 
chemical potential, using the PNJL and EPNJL models. 
In the region of $\Theta =\pi$ and $\mu_{\rm I}>m_\pi /2$, 
parity symmetry ($P$) and charge-conjugation symmetry 
($\tilde{C}$) are spontaneously broken by the pion-superfluidity 
and RW transitions, respectively. 
The  chiral and deconfinement crossovers at zero isospin and quark 
chemical potentials are remnants of these exact phase transitions, 
i.e. the $P$ restoration and the $\tilde{C}$ violation, respectively. 
The chiral and deconfinement crossovers 
almost coincide when $\Delta T=T_P-T_{\tilde{C}}$ is small. 
The interplay between the chiral and deconfinement crossovers at zero 
isospin and quark chemical potential is thus determined by 
the correlation between the $P$ restoration and the $\tilde{C}$ violation at $\mu_{\rm I}>m_\pi /2$ and $\Theta =\pi$.  

At $\mu_{\rm I} \ga 4 m_{\pi}$, the $P$ restoration becomes first-order and 
hence the $P$ restoration and the $\tilde{C}$ violation 
take place simultaneously. The $P$ restoration and the $\tilde{C}$ violation 
are strongly correlated with each other in this region.
The correlation is weakened as $\mu_{\rm I}$ decreases from $4 m_{\pi}$. 
At $\mu_{\rm I} \approx m_{\pi}\sim 2m_\pi$, 
the critical temperature of the $P$ restoration is larger 
than that of the $\tilde{C}$ violation by $10$~MeV 
in the case of the EPNJL model and by $60$~MeV in the case of the PNJL model. 
As a consequence of this property, at zero isospin and quark chemical 
potentials, the pseudocritical temperature of the chiral transition is larger 
than that of the deconfinement transition by only a few MeV 
in the case of the EPNJL model and by $50$~MeV in the case of the PNJL model. 
The correlation between the $P$ restoration and the $\tilde{C}$ violation 
at $\mu_{\rm I} \approx m_{\pi}\sim 2m_\pi$ and $\Theta=\pi$ is thus the key to 
determining the interplay between the chiral 
restoration and the deconfinement transition 
at zero isospin and quark chemical potentials.

The theoretical prediction at $\Theta=\pi$ and finite $\mu_{\rm I}$  
can be checked by LQCD, since LQCD has no sign problem there. 
Furthermore, the $P$ restoration and the $\tilde{C}$ violation are exact 
phase transitions and hence we can define the order parameters exactly. 
In future, we can thus determine the critical temperatures of the exact phase 
transitions without any ambiguity by using LQCD. 
LQCD analyses along this line may give an insight into 
the coincidence problem 
between the chiral restoration and the deconfinement transition 
at zero isospin and quark number chemical potentials. 
Furthermore, we may get knowledge of the phase diagram and the equation of state at real $\mu_{\rm I}$ and real $\mu_{\rm q}$ by the extrapolation from real $\mu_{\rm I}$ and imaginary $\mu_{\rm q}$. 

It may be also interesting to study the QCD phase diagram 
in the region where all of $\Theta$, $\mu_{\rm I}$, $\theta$ and $T$ 
are finite. For $\Theta=\mu_{\rm I}=\theta=0$ and small or zero $T$, 
only $\sigma$ is finite. 
For large $\mu_{\rm I}$ and low $T$, $\pi$ is the largest condensate 
instead of $\sigma$. At large $\theta$ and low $T$, meanwhile, 
$\eta$ is the largest one. 
For large values of $\mu_{\rm I}$ and $\theta$, it was shown 
by the NJL model~\cite{Boer} that, instead of $\pi$ or $\eta$, the charged-scalar and isovector 
condensate $a^\pm_0$ can condense at low $T$.  
 In this case, the $U_{\rm I_3}(1)$ symmetry under the transformation induced by the isospin generator $\tau_3$ is spontaneously broken, whereas 
parity symmetry is not broken. 
In principle, one can discuss the coincidence 
between the $U_{\rm I_3}(1)$ breaking and the RW  transition 
by using the PNJL model. 
The calculation, however, is quite time-consuming. 
This analysis is then an interesting future work. 

\bigskip

\noindent
\begin{acknowledgments}
Authors thank A. Nakamura, T. Inagaki, M. Matsuzaki, T. Saito, S. Nakamura, K. Morita, K. Nagata and K. Kashiwa for useful discussions and comments. 
M.K. and H. K. also thank M. Imachi, H. Yoneyama, H. Aoki, M. Tachibana and K. Fukuda for useful discussions and comments. 
T.S. and Y. S. are supported by JSPS. 
This calculation was partially carried out on SX-8 at Research Center for Nuclear Physics, Osaka University. 
\end{acknowledgments}


\end{document}